\documentclass[runningheads]{llncs}

\usepackage[misc,geometry]{ifsym}
\usepackage{graphicx}
\usepackage{booktabs}
\usepackage{caption}
\usepackage{multirow}
\usepackage{subcaption}
\usepackage{enumitem}
\usepackage{hyperref}

\newcommand{\bftab}{\fontseries{b}\selectfont}

\usepackage{xcolor}
\definecolor{boxgrey}{HTML}{F3F3F3}

\newcommand{\hlbox}[2]{
  \begin{center}
    \fcolorbox{white}{boxgrey}{
      \parbox{.9\columnwidth}{\noindent \textbf{#1}. \textit{#2}}
    }
  \end{center}
}

\begin{document}

\title{Consumer Fairness in Recommender Systems: Contextualizing Definitions and Mitigations}
\titlerunning{Consumer Fairness in Recommender Systems}

\author{Ludovico Boratto%\orcidID{0000-0002-6053-3015} 
\and Gianni Fenu%\orcidID{0000-0003-4668-2476} 
\and Mirko Marras%\orcidID{0000-0003-1989-6057}\textsuperscript{(\Letter)}  
\and Giacomo Medda}%\orcidID{0000-0002-1300-1876}} 

\authorrunning{L. Boratto et al.}

\institute{Dept. of Mathematics and Computer Science, University of Cagliari, Cagliari, Italy  \\ 
\email{\{ludovico.boratto, mirko.marras\}@acm.org, \{fenu, giacomo.medda\}@unica.it}}

\maketitle           

\begin{abstract}
Enabling non-discrimination for end-users of recommender systems by introducing consumer fairness is a key problem, widely studied in both academia and industry. 
Current research has led to a variety of notions, metrics, and unfairness mitigation procedures. 
The evaluation of each procedure has been heterogeneous and limited to a mere comparison with models not accounting for fairness. 
It is hence hard to contextualize the impact of each mitigation procedure w.r.t. the others.
In this paper, we conduct a systematic analysis of mitigation procedures against consumer unfairness in rating prediction and top-n recommendation tasks. 
To this end, we collected 15 procedures proposed in recent top-tier conferences and journals. Only 8 of them could be reproduced. 
Under a common evaluation protocol, based on two public data sets, we then studied the extent to which recommendation utility and consumer fairness are impacted by these procedures, the interplay between two primary fairness notions based on equity and independence, and the demographic groups harmed by the disparate impact. 
Our study finally highlights open challenges and future directions in this field.
The source code is available at \url{https://github.com/jackmedda/C-Fairness-RecSys}. 
\keywords{Recommender Systems \and Fairness \and Bias \and Consumers.}
\end{abstract}

\section{Introduction} \label{sec:intro}

Recommender systems help us make decisions, from selecting books to choosing friends~\cite{DBLP:reference/sp/2015rsh}. 
Their wide adoption has spurred investigations into possibly unfair practices in the systems' mechanisms~\cite{DBLP:journals/corr/abs-2010-03240,DBLP:journals/corr/abs-2105-05779,DBLP:journals/ipm/DeldjooBN21,marras2021equality,DBLP:journals/umuai/BorattoFM21}. 
Fairness is a concept of \emph{non-discrimination} on the basis of the membership to \emph{protected groups}, identified by a \emph{protected feature}, e.g., gender and age in anti-discrimination legislation\footnote{Please refer to Art. 21 of the EU Charter of Fundamental Rights, Art. 14 of European Convention on Human Rights, Art. 18-25 of the Treaty on the Functioning of EU.}. 
\emph{Group fairness} avoids the discrimination of a given group, assessed as the absence of a \emph{disparate impact} in the outcomes generated for them~\cite{DBLP:journals/csur/MehrabiMSLG21}. 
Despite involving different stakeholders (e.g., providers and sellers), fairness in recommender systems may particularly affect those who receive the recommendations (\emph{consumers})~\cite{DBLP:journals/umuai/AbdollahpouriAB20}. 
Hence, group consumer fairness should account for no disparate impact of recommendations on protected groups of consumers. 
Providing guarantees on this property is a key strategic objective for the responsible advancement of the field.  

As fairness is an abstract concept, an abundance of consumer fairness notions have been proposed, along with algorithmic procedures for mitigating unfairness in recommendations according to the proposed notions. 
Examples of mitigation procedures have been applied in \emph{pre-processing}~\cite{DBLP:conf/fat/EkstrandTAEAMP18}, by transforming the input data, \emph{in-processing}~\cite{DBLP:conf/fat/KamishimaAAS18,DBLP:conf/fat/BurkeSO18,DBLP:conf/ecmlpkdd/Frisch21,DBLP:conf/www/WuCSHWW21}, by constraining the training process of state-of-the-art models, and \emph{post-processing}~\cite{DBLP:conf/www/LiCFGZ21,DBLP:conf/wsdm/RastegarpanahGC19,DBLP:journals/ipm/AshokanH21}, by ranking again the originally recommended items.
Moreover, the evaluation protocol adopted to assess their impact has been often \emph{heterogeneous} (e.g., different data sets, train-test splits) and limited to showing that the proposed mitigation is \emph{better than doing nothing}, making the landscape convoluted. 
To shape recommender systems that account for consumer fairness, we need a common understanding and practical benchmarks on how and when each procedure can be used in comparison to the others. As a response, with this research work, we address \emph{three research questions}:

\vspace{-1mm}
\begin{enumerate}[label=\textbf{RQ\arabic*},leftmargin=10mm]
\item \emph{Is recommendation utility affected by the mitigation procedures?} 
\item \emph{Do the selected mitigation procedures reduce the unfairness estimates? }
\item \emph{Is disparate impact systematically harming the minority group?}
\end{enumerate}
\vspace{-1mm}

To answer these questions, in a first step (Section~\ref{sec:researchmethod}), we conducted a systematic study on algorithmic procedures for mitigating consumer unfairness in rating prediction or top-n recommendation tasks. 
To this end, we scanned the proceedings of top-tier conferences and journals, identifying \emph{15 relevant papers}. 
We tried to reproduce the procedures reported in the paper in case the source code was made available by the authors (\emph{only 8 papers}). 
Our first contribution is hence an assessment of the reproducibility level of mitigations in the area.

In a second step (Section~\ref{sec:results}), we defined a common evaluation protocol, including two public data sets (MovieLens 1M; LastFM 1K), two sensitive attributes (gender; age) and two fairness notions (equity; independence); we evaluated the recommendation models reported in the papers, with/out the proposed mitigation procedure, under this common protocol. 
Our results revealed that, the mitigation procedures did not consistently reduce the utility of the recommendations (\emph{RQ1}). 
We however found that only a minor subset of procedures substantially reduce unfairness, and rarely for the two fairness notions at the same time (\emph{RQ2}). 
Moreover, disparate impact does not always harm minority groups (\emph{RQ3}). 
Our second contribution lies in evaluating mitigation procedures under a common protocol and identifying challenges in the area (Section~\ref{sec:discussion}). 

\section{Research Methodology} \label{sec:researchmethod}
\begin{figure}[!t]
\includegraphics[width=\textwidth]{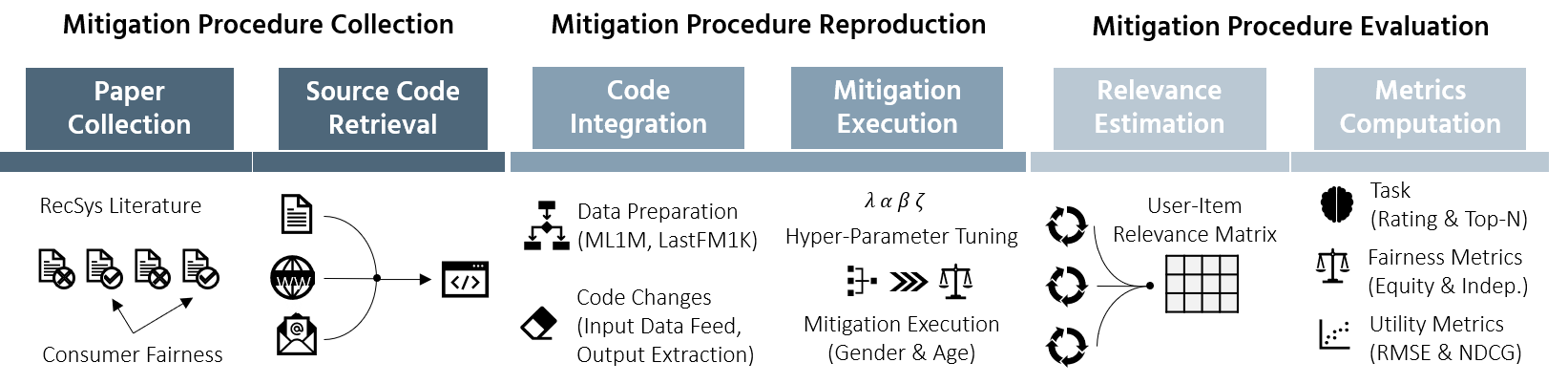}
\vspace{-6mm}
\caption{\footnotesize \textbf{Method}. We systematically collected papers and retrieved their source code. We processed the data sets used in our evaluation protocol, formatted them as per each mitigation requirements, and made the format of the mitigation results uniform. We trained the recommendation models included in the original papers, with/out mitigation, and computed fairness and utility metrics for the target recommendation task.} 
\vspace{-4mm}
\label{fig:schema}
\end{figure}

\vspace{-1mm}
In this section, we describe the collection process for mitigation procedures, the steps for their reproduction, and the common evaluation protocol (Figure \ref{fig:schema}).      

\subsection{Mitigation Procedures Collection} \label{subsec:collection}
To collect existing mitigation procedures against consumer fairness, we systematically scanned the recent proceedings of top-tier Information Retrieval conferences and workshops, namely CIKM, ECIR, ECML-PKDD, FAccT, KDD, RecSys, SIGIR, WSDM, WWW, and journals edited by top-tier publishers, namely ACM, Elsevier, IEEE, and Springer. The keywords for our manual research were composed by a technical term, ``\emph{Recommender System}'' or ``\emph{Recommendation}'', and a non-technical term, ``\emph{Consumer Fairness}'' or ``\emph{User Fairness}''. We marked a paper to be relevant if (a) it focused on recommender systems, (b) it proposed a mitigation procedure, and (c) that procedure targeted the end users receiving the recommendations. Papers on other domains, e.g., non-personalized rankings, other stakeholders, e.g., providers only, and on pure conceptualization only, e.g., proposing a fairness notion without any mitigation, were excluded. Papers addressing both consumer and provider fairness were included, since they also target the end users. Finally, \emph{15 relevant papers} were considered in our study.

We then attempted to reproduce the mitigation procedure proposed in each relevant paper, relying as much as possible on the source code provided by the authors themselves. 
We hence tried to obtain the source code for each relevant paper, by searching for the link into the paper, browsing for the official repository on the Web
, and sending an e-mail to the authors as a last resort. We considered a mitigation procedure to be reproducible if a working version of the source code was obtained, and required minimal changes to accept another data set and extract the final recommendations. Otherwise, we considered a paper to be non-reproducible given our reproduction approach. We also considered works to be non-reproducible when the source code was obtained but included only a skeleton version of the procedure with many parts and details missing. At the end, \emph{8 out of 15 relevant papers} could be reproduced with a reasonable effort. 

In Table~\ref{tab:repr_papers}, for each reproducible paper, we identified the recommendation task (RP : Rating Prediction; TR : Top-N Recommendation), the notion of consumer fairness (\texttt{EQ} : equity of the error/utility score across demographic groups; \texttt{IND} : independence of the predicted relevance scores or recommendations from the demographic group), the consumers' grouping (\texttt{G} : Gender, \texttt{A} : Age, \texttt{O} : Occupation, \texttt{B} : Behavioral), the mitigation type (\texttt{PRE}-, \texttt{IN}- or \texttt{POST}-Processing), the evaluation data sets (\texttt{ML} : MovieLens 1M or 10M, \texttt{LFM} : LastFM 1K or 360K, \texttt{AM}: Amazon, \texttt{SS}: Sushi, \texttt{SY}: Synthetic), the utility/accuracy metrics (\texttt{NDCG} : Normalized Discounted Cumulative Gain; \texttt{F1} : F1 Score; \texttt{AUC}: Area Under Curve; \texttt{MRR} : Mean Reciprocal Rank; \texttt{RMSE} : Root Mean-Square Error; \texttt{MAE} : Mean Absolute Error), and fairness metrics (\texttt{EPS} : $\epsilon$-fairness; \texttt{CHI} : Chi-Square Test; \texttt{KS} : Kolmogorov-Smirnov Test; \texttt{GEI} : Generalized Entropy Index; \texttt{TI} : Theil Index; \texttt{DP} : Demographic Parity; \texttt{EP}: Equal Opportunity;  \texttt{CES} : Category Equity Score; \texttt{GLV}: Group Loss Variance). The reproducibility ratio was of 53\% (8/15) in total: 50\% (4/8) for top-n recommendation and 57\% (4/7) for rating prediction. We identified \cite{fairrec,multifr,causalnotion,biasdisparity} and \cite{deepfair,bandits,fairhybrid} as non-reproducible procedures according to our criteria for top-n recommendation and rating prediction, respectively. 

\begin{table}[!t]
\centering
\caption{\footnotesize The considered reproducible mitigation procedures for consumer fairness.}
\label{tab:repr_papers}
\resizebox{1.0\textwidth}{!}{
\begin{tabular}{l l l | l l l | l l l}
\toprule
\multirow{2}{*}{\textbf{Task}} & \multirow{2}{*}{\textbf{Paper}} & \multirow{2}{*}{\textbf{Year}} & \multicolumn{3}{l}{\textbf{Mitigation}}      & \multicolumn{2}{|l}{\textbf{Evaluation}} \\
                                &                            &    & Notion & Groups & Type & Data Sets         & Utility Metrics & Fairness Metrics         \\
\midrule
\multirow{3}{*}{TR} & Burke et al.~\cite{DBLP:conf/fat/BurkeSO18} & 2018           &         EQ         &         G          &        IN     &        ML          &    NDCG & CES        \\
& Frisch et al.~\cite{DBLP:conf/ecmlpkdd/Frisch21} & 2021                           &         IND          &         G-A          &        IN     &        ML          &     NDCG & EPS-CHI        \\
& Li et al.~\cite{DBLP:conf/www/LiCFGZ21} & 2021                               &         EQ          &         B         &        POST   &        AM          &     NDCG-F1  &  DP      \\
\midrule
TR + RP & Ekstrand et al.~\cite{DBLP:conf/fat/EkstrandTAEAMP18} & 2018                         &         EQ          &         G          &     PRE       &        ML-LFM      &    NDCG-MRR & DP       \\
\midrule
\multirow{4}{*}{RP} & Kamishima et al.~\cite{DBLP:conf/fat/KamishimaAAS18} & 2018   &         IND         &         G-A        &        IN     &        ML-SS       &   MAE & KS    \\
& Rastegarpanah et al.~\cite{DBLP:conf/wsdm/RastegarpanahGC19} & 2019                    &         EQ          &         B           &        POST   &        ML       &   RMSE & GLV         \\
& Ashokan \& Haas~\cite{DBLP:journals/ipm/AshokanH21} & 2021                      &         EQ          &         G          &        POST   &        ML-SY       &     RMSE-MAE & %YH-
GEI-TI         \\
& Wu et al.~\cite{DBLP:conf/www/WuCSHWW21} & 2021                                    &         IND          &         G-A-O      &        IN     &        ML-LFM      &     RMSE & AUC-F1         \\
\bottomrule  
\end{tabular}
}
\vspace{-5mm}
\end{table}

\subsection{Mitigation Procedures Reproduction} \label{subsec:repralg}
For each reproducible paper, we delve into the core idea and the characteristics reported in Table~\ref{tab:repr_papers}. Our source code includes a directory for each paper, documented with the changes on the original code and the steps to get our results. 

\vspace{1mm} \noindent \textbf{Burke et al.}~\cite{DBLP:conf/fat/BurkeSO18} proposed to generate recommendations for a user from a neighborhood having an equal number of peers from each group, to reduce unfairness. SLIM, a collaborative filtering method, was extended with a regularization aimed to achieve balance between protected and non-protected neighbors. Fairness was measured with a variant of what is known in statistics as risk ratio; this score is less (greater) than 1 when the protected group is recommended fewer (more) movies of the desired genre, on average (1 means perfect equity). Recommendation utility was measured via NDCG@10. An evaluation on ML 1M (5-fold cross-validation, no train-test split specified) showed that the treated models led to an equity score closer to 1 than the original models. The source code was not included in the paper, but shared by the authors during a scientific tutorial \cite{DBLP:conf/fat/BurkeMS20}.  

\vspace{1mm} \noindent \textbf{Frisch et al.}~\cite{DBLP:conf/ecmlpkdd/Frisch21} aimed at producing fair recommendations using a co-clustering of users and items that respects statistical parity w.r.t. some sensitive attributes. To this end, the authors introduced a co-clustering model based on the Latent Block Model (LBM), that relies on an ordinal regression model taking the sensitive attributes as inputs. Fairness was measured by monitoring that, for any two items, the proportion of users with the same preference was similar across demographic groups. NDCG monitored the recommendation utility. An evaluation on ML 1M showed that their procedure led to lower unfairness. No source code was linked in the paper; we contacted the authors to get a copy of it. 

\vspace{1mm} \noindent \textbf{Li et al.}~\cite{DBLP:conf/www/LiCFGZ21} investigated consumer unfairness across user groups based on the level of activity in the platform (more or less active). As a mitigation, the authors adopted a re-ranking method, whose objective function was to select items out of the baseline top-n list of each user so that the overall recommendation utility could be maximized, constrained to the fact that the model should minimize the difference in average recommendation performance between the groups of users. F1@10 and NDCG@10 were used to assess recommendation utility. The difference in NDCG between the groups estimated the unfairness of the model. An evaluation on Amazon data sets showed that their procedure could reduce unfairness between groups significantly, and also improve the overall recommendation utility. The original source code in the paper included only the re-ranking method. We contacted the authors for the complete source code,  which was provided in a public repository (\url{https://github.com/rutgerswiselab/NLR}).

\vspace{1mm} \noindent \textbf{Ekstrand et al.}~\cite{DBLP:conf/fat/EkstrandTAEAMP18} re-sampled user interactions (random sampling without replacement), such that the representation of user interactions across groups in the training set was balanced, and re-trained the recommendation models with the balanced training set. Recommendation utility was measured with NDCG@10, and fairness was assessed by visually comparing the averaged NDCG scores for the different demographic groups. An evaluation on ML 1M and LFM 1K and 360K showed that their re-sampling procedure led to unfairness mitigation for gender groups. The link to the source code was reported in the paper, pointing to a publicly available archive stored in the authors' university website (\url{https://scholarworks.boisestate.edu/cs\_scripts/4/}).

\vspace{1mm} \noindent \textbf{Kamishima et al.}~\cite{DBLP:conf/fat/KamishimaAAS18} delved into the concept of recommendation independence,  achieved when a recommendation outcome (predicted ratings) is statistically independent from a specified sensitive attribute. The mitigation consisted of optimizing a recommendation model by minimizing the dissimilarity between true ratings and predicted ratings and jointly maximizing the degree of independence between the predicted ratings and sensitive labels. Prediction errors were measured by the MAE. Independence was checked by measuring the equality of the predicted rating distributions between groups (Kolmogorov-Smirnov test; a smaller KS indicates that predicted ratings are more independent). An evaluation on ML 1M, Flixster, and Sushi, and three independence terms (mean-m, bdist-m, and mi-normal), showed that the sensitive information could be removed at the cost of a small loss in MAE. 
The source code linked in the paper included two complementary repositories (\url{https://github.com/tkamishima/kamrecsys}) and (\url{https://github.com/tkamishima/kamiers}). 

\vspace{1mm} \noindent \textbf{Rastegarpanah et al.}~\cite{DBLP:conf/wsdm/RastegarpanahGC19} investigated whether augmenting the training input with additional data can improve the fairness of the resulting predictions. Given a pretrained matrix factorization model, the mitigation required to add fake users who provided ratings on existing items to the training set; the fake users’ ratings were chosen to improve the fairness of the final model for the real users. Recommendation utility was measured via RMSE. Fairness was measured through the variance of the loss across demographic groups, with the loss being the mean squared estimation error over all ratings of users in the group. An evaluation on ML 1M (no train-test split specified) showed that their mitigation could efficiently improve fairness of the considered recommender systems. The source code was found in a public repository  (\url{https://github.com/rastegarpanah/antidote-data-framework}), whose link was sent to us by the authors.

\vspace{1mm} \noindent \textbf{Ashokan \& Haas}~\cite{DBLP:journals/ipm/AshokanH21} adjusted the relevance scores predicted by the original model such that a given fairness metric increased. The authors experimented with value-based fairness (given a user, the difference in predicted and actual ratings in the training set for the group the user belongs to was added to the predicted ratings of the user) and parity-based fairness (the overall difference between predicted ratings for two groups on the training set was added to the predicted rating of a user for an item in the test set). Recommendation utility was measured via RMSE and MAE. Fairness was measured, among others, via the Generalized Entropy and the Theil indexes, which estimate inequality of errors across users. An evaluation on the ALS and ItemKNN recommendation models, trained on a synthetic data set and on ML 1M (5-fold cross-validation), showed that increasing fairness can even lead to lower RMSE and MAE in certain cases. No source code was linked in the paper; the authors sent it to us by e-mail.

\vspace{1mm} \noindent \textbf{Wu et al.}~\cite{DBLP:conf/www/WuCSHWW21} focused on mitigating unfairness in latent factor models. To this end, their procedure took the user and item embeddings from the original recommendation model as input and learned a filter space where any sensitive information was obfuscated and recommendation utility was preserved. The filters were learnt through a graph-based adversarial training process, where a discriminator tried to predict the sensitive label, and the filters were trained to remove sensitive information exposed in the supporting graph structure. RMSE measured recommendation utility. Fairness was monitored by checking the performance in terms of AUC (binary attributes) and F1 (multi-class attributes) of a classifier that predicts the sensitive attribute, given the user embedding (smaller values denote better fairness). An evaluation on ML 1M (training and test ratio of 9:1) and LFM 360K (training, validation, test ratio of 7:1:2) showed that fairness could be improved without significantly impacting on recommendation utility. The source code linked in the paper omitted important components. The authors provided us with an updated public repository (\url{https://github.com/newlei/LR-GCCF}).

\subsection{Mitigation Procedures Evaluation} \label{subsec:evalmethod}

To ensure evaluation consistency and uniformity across mitigation procedures, given the heterogeneity of the original experimental evaluations, we mixed replication and reproduction~\cite{acmbadging,DBLP:conf/recsys/DacremaCJ19}. For readability, we use the term ``\emph{reproducibility}''. So, we used the source code provided by the original authors to run their models and mitigation procedures, and our own artifacts (data and source code) to (a) pre-process the input data sets as per their requirements and (b) compute evaluation metrics based on the relevance scores or recommendations they returned. 

\vspace{1mm} \noindent \textbf{Data Sets}. The assessment of consumer fairness is challenging due to the lack of public data sets with ratings and \emph{sensitive attributes} of the consumers. In our analysis, we considered all the public data sets that (a) were used in at least one reproduced paper, (b) reported at least one sensitive attribute, and (c) included enough ratings to reasonably train a recommender system ($\ge$ 200,000 ratings). We hence evaluated the reproduced mitigation procedures on two public data sets on the movies and music domains (Table~\ref{tab:repr_datasets}). Each data set was downloaded from the original website and pre-processed according to our common evaluation protocol, in response also to some limitations of the reproduced mitigations. For instance, given that the existing mitigation procedures are often tailored to binary groups only, we grouped users in two groups in case of data sets with multi-class sensitive attributes (\emph{while attributes like gender and age are by no means a binary construct, what we are considering is a binary feature}). 

Gender labels were already binary in ML 1M. We binarized age labels, organized in seven age ranges, such that the two groups included consecutive age ranges and had the most similar representation possible. For LFM 1K, we considered only users reporting both their gender and age and filtered those with wrong ages ($\le 0$ or $\ge 125$). Interactions of a user for the same artist were aggregated, using the number of plays of a user for an artist as a proxy of the rating. We filtered users interacting with less than 20 artists (as in ML 1M), and ratings were log-normalized and scaled in $[1,5]$. Gender labels were already binary. We binarized age labels (integer) with the same criteria used in ML 1M.   

\begin{table}[!t]
\centering
\caption{\footnotesize The data sets with consumer's sensitive attributes included in our study.}
\label{tab:repr_datasets}
\resizebox{1.0\textwidth}{!}{
\begin{tabular}{llll|l}
\toprule
\textbf{Data Set} & \textbf{\#Users} & \textbf{\#Items} & \textbf{\#Ratings} & \textbf{Sensitive Attributes} \\
\midrule
ML 1M \cite{DBLP:journals/tiis/HarperK16}                &        6,040          &      3,952            &      1,000,209              &  Gender (M : 71.7\%; F : 28.3\%) Age ( $< 35$ : 56.6\%; $\ge 35$ : 43.4\%) \\
LFM 1K \cite{DBLP:books/daglib/0025137}                &        268          &     51,609             &        200,586            &    Gender (M : 57.8\%; F : 42.2\%) Age ( $< 25$ : 57.8\%; $\ge 25$ : 42.2\%)                   \\
\bottomrule                          
\end{tabular}
}
\vspace{-6mm}
\end{table}

\vspace{1mm} \noindent \textbf{Protocol}. Each reproduced paper applied the corresponding mitigation procedure to a set of state-of-the-art recommendation models, which was quite heterogeneous across papers due to authors' arbitrary choices or the focus on a specific type of model. These models covered several families, including non-personalized (\texttt{TopPopular}~\cite{DBLP:conf/fat/EkstrandTAEAMP18} and \texttt{AvgRating}~\cite{DBLP:conf/fat/EkstrandTAEAMP18}), memory (\texttt{ItemKNN}~\cite{DBLP:conf/fat/EkstrandTAEAMP18,DBLP:journals/ipm/AshokanH21}, \texttt{UserKNN}~\cite{DBLP:conf/fat/EkstrandTAEAMP18}), matrix factorization (\texttt{BiasedMF}~\cite{DBLP:conf/www/LiCFGZ21,DBLP:journals/ipm/AshokanH21}, \texttt{PMF}~\cite{DBLP:conf/www/LiCFGZ21,DBLP:conf/fat/KamishimaAAS18,DBLP:conf/www/WuCSHWW21}, \texttt{FunkSVD}~\cite{DBLP:conf/fat/EkstrandTAEAMP18}), learning-to-rank (\texttt{NCF}~\cite{DBLP:conf/www/LiCFGZ21}, \texttt{LBM}~\cite{DBLP:conf/ecmlpkdd/Frisch21}, \texttt{SLIM-U}~\cite{DBLP:conf/fat/BurkeSO18}, \texttt{ALS}~\cite{DBLP:conf/wsdm/RastegarpanahGC19}, \texttt{LMaFit}~\cite{DBLP:conf/wsdm/RastegarpanahGC19}), graph (\texttt{GCN}~\cite{DBLP:conf/www/WuCSHWW21}), and session-based (\texttt{STAMP}~\cite{DBLP:conf/www/LiCFGZ21}). In line with our reproduction approach, we applied a given mitigation on the same models considered by the original authors\footnote{Though some procedures might be applied across models, their transfer often requires arbitrary design choices and core changes that mine our rigorous reproduction.}.   

Specifically, given a data set, a sensitive attribute, and a reproducible paper, we considered the following evaluation protocol. We first performed a train-test split per user, with 20\% of the interactions (the most recent if a timestamp was available, randomly selected otherwise) being in the test set and the remaining interactions being in the train set. In case a validation set was needed for best model selection, 10\% of interactions (selected in the same way) of each user from the train set were considered as a validation set and the other ones included in the final train set. To fit with the original source code, the format of the considered sets and the sensitive attribute's labels per user were adapted. No changes on the source code specific for the mitigation procedure were applied. 

Using the prepared sets and an appropriate hyper-parameters grid, we ran a grid search for each recommendation model, with and without mitigation. For each paper, our source code includes the scripts to format a data set as per the original source code requirements and to compute evaluation metrics as well as the details of models hyper-parameter tuning. For each setup, we obtained the predicted relevance scores and the recommendations, and computed utility and fairness metrics. Utility metrics included NDCG for top-n recommendation (using binary relevances) and RMSE for rating prediction, selected due to their popularity (see Table \ref{tab:repr_papers}). Consumer fairness metrics monitored equity through Demographic Parity (DP), computed as the difference on utility for the corresponding task between groups, and independence through Kolmogorov-Smirnov (KS), computed on predicted relevance scores, covering two well-known perspectives and steps of the pipeline. Mainly due to space constraints, we left analyses on other fairness notions and implementations of the same fairness notions as a future work. 
Experiments ran on a Ryzen7 machine with 32 GB RAM.      

\vspace{-2mm}
\section{Experimental Results} \label{sec:results}
We now analyze the extent to which the mitigation procedures impact on recommendation utility (\emph{RQ1}), reduce unfairness (\emph{RQ2}), and possibly affect groups differently (\emph{RQ3}). To this end, we report recommendation utility and fairness scores obtained under the above evaluation protocol, for TR (Table~\ref{tab:ndcg_gender}, gender; Table \ref{tab:ndcg_age}, age) and RP tasks (Table~\ref{tab:rmse_gender}, gender; Table \ref{tab:rmse_age}, age). DP was tested for statistical significance via a Mann-Whitney test. For KS, we used its own score. Note that * and $\wedge$ meant significance at p-values $0.05$ and $0.01$, respectively.  
\vspace{-3mm}

\begin{table}[!b]
\centering
\vspace{-7mm}
\caption{\footnotesize Top-n recommendation (TR) considering \emph{gender} groups.}
\label{tab:ndcg_gender}
\resizebox{1.0\textwidth}{!}{
\begin{tabular}{ll|rrrrrr|rrrrrr}
\toprule
          &       & \multicolumn{6}{c|}{ML 1M} & \multicolumn{6}{c}{LFM 1K} \\
          &       & \multicolumn{2}{c}{NDCG $\uparrow$} & \multicolumn{2}{c}{DP $\downarrow$} & \multicolumn{2}{c|}{KS $\downarrow$} & \multicolumn{2}{c}{NDCG $\uparrow$} & \multicolumn{2}{c}{DP $\downarrow$} & \multicolumn{2}{c}{KS $\downarrow$} \\
          Paper & Model &           \multicolumn{1}{c}{Base} &           \multicolumn{1}{c}{Mit} &                      \multicolumn{1}{c}{Base} &                      \multicolumn{1}{c}{Mit} &                      \multicolumn{1}{c}{Base} &                      \multicolumn{1}{c|}{Mit} &           \multicolumn{1}{c}{Base} &           \multicolumn{1}{c}{Mit} &                       \multicolumn{1}{c}{Base} &                       \multicolumn{1}{c}{Mit} &                      \multicolumn{1}{c}{Base} &                      \multicolumn{1}{c}{Mit} \\
\midrule
Burke et al. & SLIM-U &        0.084 &  0.084 &  {\scriptsize \^{}}0.022 &  {\scriptsize \^{}}0.028 &  {\scriptsize \^{}}0.032 &  {\scriptsize \^{}}0.115 &         0.348 &  0.301 &  {\scriptsize \^{}}-0.128 &  {\scriptsize \^{}}0.072 &  {\scriptsize \^{}}0.010 &  {\scriptsize \^{}}0.142 \\
Frisch et al. & LBM &         0.044 &  0.021 &  {\scriptsize \^{}}0.006 &  {\scriptsize \^{}}0.004 &  {\scriptsize \^{}}0.013 &  {\scriptsize \^{}}0.025 &         0.144 &  0.212 &  {\scriptsize *}-0.035 &  {\scriptsize *}-0.058 &  {\scriptsize \^{}}0.120 &  {\scriptsize \^{}}0.126 \\
Li et al. & BiasedMF &         0.112 &         0.112 &  {\scriptsize \^{}}0.016 &  {\scriptsize \^{}}0.013 &  {\scriptsize \^{}}0.033 & \bftab {\scriptsize \^{}}0.006 &         0.246 &   0.245 &  {\scriptsize \^{}}-0.076 &  {\scriptsize *}-0.049 &  {\scriptsize \^{}}0.026 & \bftab {\scriptsize \^{}}0.001 \\
          & NCF &         0.120 &         0.120 &  {\scriptsize \^{}}0.018 &  {\scriptsize \^{}}0.015 &  {\scriptsize \^{}}0.024 & \bftab {\scriptsize \^{}}0.006 &         0.204 &         0.202 &                    -0.046 &                 -0.023 &  {\scriptsize \^{}}0.017 & \bftab {\scriptsize \^{}}0.001 \\
          & PMF &         0.123 &   0.123 &  {\scriptsize \^{}}0.020 &  {\scriptsize \^{}}0.015 &  {\scriptsize \^{}}0.026 & \bftab {\scriptsize \^{}}0.006 &         0.163 &         0.164 &  {\scriptsize \^{}}-0.069 &  {\scriptsize *}-0.049 &  {\scriptsize \^{}}0.035 & \bftab {\scriptsize \^{}}0.001 \\
          & STAMP &         0.068 &         0.067 &  {\scriptsize \^{}}0.013 &  {\scriptsize \^{}}0.009 & \bftab {\scriptsize \^{}}0.007 & \bftab {\scriptsize \^{}}0.006 &         0.110 &         0.110 &              -0.024 &           -0.018 &  {\scriptsize \^{}}0.002 & \bftab {\scriptsize \^{}}0.001 \\
Ekstrand et al. & FunkSVD &         0.018 &         0.015 & \bftab {\scriptsize \^{}}0.004 &       \bftab       0.002 &  {\scriptsize \^{}}0.027 &  {\scriptsize \^{}}0.018 &         0.010 &         0.013 &      \bftab        -0.006 &             \bftab -0.003 &  {\scriptsize \^{}}0.107 &  {\scriptsize \^{}}0.119 \\
          & ItemKNN &   \bftab 0.140 &  \bftab 0.134 &  {\scriptsize \^{}}0.038 &  {\scriptsize \^{}}0.030 &  {\scriptsize \^{}}0.030 &  {\scriptsize \^{}}0.031 &         0.287 &         0.286 &  {\scriptsize \^{}}-0.127 &     {\scriptsize *}-0.116 &  {\scriptsize \^{}}0.019 &  {\scriptsize \^{}}0.022 \\
                
          & TopPopular &         0.110 &         0.104 &  {\scriptsize \^{}}0.035 &  {\scriptsize \^{}}0.030 & \bftab {\scriptsize \^{}}0.007 &  {\scriptsize \^{}}0.007 &         0.312 &         0.321 &     {\scriptsize *}-0.085 &     {\scriptsize *}-0.102 &  \bftab {\scriptsize \^{}}0.001 &  {\scriptsize \^{}}0.002 \\
          & UserKNN &         0.137 &         0.131 &  {\scriptsize \^{}}0.031 &  {\scriptsize \^{}}0.024 &  {\scriptsize \^{}}0.074 &  {\scriptsize \^{}}0.052 &  \bftab 0.406 & \bftab  0.411 &  {\scriptsize \^{}}-0.110 &  {\scriptsize \^{}}-0.106 &  {\scriptsize \^{}}0.067 &  {\scriptsize \^{}}0.067 \\
\bottomrule
\end{tabular}
}
\end{table}

\begin{table}[!t]
\centering
\caption{\footnotesize Top-n recommendation (TR) considering \emph{age} groups.}
\label{tab:ndcg_age}
\resizebox{1.0\textwidth}{!}{
\begin{tabular}{ll|rrrrrr|rrrrrr}
\toprule
          &       & \multicolumn{6}{c|}{ML 1M} & \multicolumn{6}{c}{LFM 1K} \\
          &       & \multicolumn{2}{c}{NDCG $\uparrow$} & \multicolumn{2}{c}{DP $\downarrow$} & \multicolumn{2}{c|}{KS $\downarrow$} & \multicolumn{2}{c}{NDCG $\uparrow$} & \multicolumn{2}{c}{DP $\downarrow$} & \multicolumn{2}{c}{KS $\downarrow$} \\
          Paper & Model &           \multicolumn{1}{c}{Base} &           \multicolumn{1}{c}{Mit} &                      \multicolumn{1}{c}{Base} &                      \multicolumn{1}{c}{Mit} &                      \multicolumn{1}{c}{Base} &                      \multicolumn{1}{c|}{Mit} &           \multicolumn{1}{c}{Base} &           \multicolumn{1}{c}{Mit} &                       \multicolumn{1}{c}{Base} &                       \multicolumn{1}{c}{Mit} &                      \multicolumn{1}{c}{Base} &                      \multicolumn{1}{c}{Mit} \\
\midrule
Burke et al. & SLIM-U &         0.084 &  0.048 &  {\scriptsize \^{}}0.022 &  {\scriptsize \^{}}0.014 &  {\scriptsize \^{}}0.009 &  {\scriptsize \^{}}0.095 &         0.348 &  0.207 &  {\scriptsize *}-0.065 &  {\scriptsize \^{}}-0.145 &  {\scriptsize \^{}}0.021 &  {\scriptsize \^{}}0.082 \\
Frisch et al. & LBM &         0.044 &   0.042 & \bftab {\scriptsize \^{}}0.005 & \bftab {\scriptsize \^{}}0.006 &  {\scriptsize \^{}}0.021 &  {\scriptsize \^{}}0.027 &         0.144 &   0.213 &   -0.011 &   -0.021 &  {\scriptsize \^{}}0.125 &  {\scriptsize \^{}}0.152 \\
Li et al. & BiasedMF &         0.112 &         0.112 &  {\scriptsize \^{}}0.018 &      {\scriptsize \^{}}0.017 &  {\scriptsize \^{}}0.042 & \bftab {\scriptsize \^{}}0.006 &         0.246 &   0.247 &                        -0.044 &     {\scriptsize *}-0.060 &  {\scriptsize \^{}}0.015 & \bftab {\scriptsize \^{}}0.005 \\
          & NCF &         0.120 &         0.120 &  {\scriptsize \^{}}0.022 &      {\scriptsize \^{}}0.019 &  {\scriptsize \^{}}0.031 & \bftab {\scriptsize \^{}}0.006 &         0.204 &         0.203 &                        -0.035 &                    -0.048 &  {\scriptsize \^{}}0.008 & \bftab {\scriptsize \^{}}0.005 \\
          & PMF &         0.123 &   0.123 &  {\scriptsize \^{}}0.027 &      {\scriptsize \^{}}0.021 &  {\scriptsize \^{}}0.027 & \bftab {\scriptsize \^{}}0.006 &         0.163 &         0.164 &                        -0.033 &  {\scriptsize \^{}}-0.044 &  {\scriptsize \^{}}0.018 & \bftab {\scriptsize \^{}}0.005 \\
          & STAMP &         0.068 &         0.068 &       \bftab       0.005 & \bftab  {\scriptsize *}0.006 & \bftab {\scriptsize \^{}}0.006 & \bftab {\scriptsize \^{}}0.006 &         0.110 &         0.110 &   {\scriptsize *}-0.030 &  {\scriptsize \^{}}-0.034 & \bftab {\scriptsize \^{}}0.005 & \bftab {\scriptsize \^{}}0.005 \\
Ekstrand et al. & FunkSVD &         0.018 &         0.016 &  {\scriptsize \^{}}0.008 & \bftab {\scriptsize \^{}}0.006 &  {\scriptsize \^{}}0.029 &  {\scriptsize \^{}}0.021 &         0.010 &         0.016 &  \bftab 0.002 & \bftab  -0.004 &  {\scriptsize \^{}}0.054 &  {\scriptsize \^{}}0.047 \\
          & ItemKNN & \bftab  0.140 & \bftab  0.138 &  {\scriptsize \^{}}0.027 &  {\scriptsize \^{}}0.024 &  {\scriptsize \^{}}0.029 &  {\scriptsize \^{}}0.033 &         0.287 &         0.269 &         0.010 &          0.020 &  {\scriptsize \^{}}0.133 &  {\scriptsize \^{}}0.118 \\
          & TopPopular &         0.110 &         0.107 &  {\scriptsize \^{}}0.038 &  {\scriptsize \^{}}0.034 &  \bftab {\scriptsize \^{}}0.006 &  \bftab {\scriptsize \^{}}0.006 &         0.312 &         0.315 &                    -0.044 &                    -0.050 &  {\scriptsize \^{}}0.006 &  {\scriptsize \^{}}0.007 \\
          & UserKNN &         0.137 &         0.137 &  {\scriptsize \^{}}0.028 &  {\scriptsize \^{}}0.023 &  {\scriptsize \^{}}0.060 &  {\scriptsize \^{}}0.051 & \bftab  0.406 & \bftab  0.397 &        -0.023 &         -0.031 &  {\scriptsize \^{}}0.036 &  {\scriptsize \^{}}0.031 \\
\bottomrule
\end{tabular}
}
\end{table}

\begin{table}[!t]
\vspace{-7mm}
\centering
\caption{\footnotesize Rating prediction (RP) considering gender groups.}
\label{tab:rmse_gender}
\resizebox{1.0\textwidth}{!}{
\begin{tabular}{ll|rrrrrr|rrrrrr}
\toprule
          &            & \multicolumn{6}{c|}{ML 1M} & \multicolumn{6}{c}{LFM 1K} \\
          &            & \multicolumn{2}{c}{RMSE $\downarrow$} & \multicolumn{2}{c}{DP $\downarrow$} & \multicolumn{2}{c|}{KS $\downarrow$} & \multicolumn{2}{c}{RMSE $\downarrow$} & \multicolumn{2}{c}{DP $\downarrow$} & \multicolumn{2}{c}{KS $\downarrow$} \\
          Paper & Model &           \multicolumn{1}{c}{Base} &           \multicolumn{1}{c}{Mit} &                           \multicolumn{1}{c}{Base} &                           \multicolumn{1}{c}{Mit} &                      \multicolumn{1}{c}{Base} &                      \multicolumn{1}{c|}{Mit} &           \multicolumn{1}{c}{Base} &           \multicolumn{1}{c}{Mit} &                      \multicolumn{1}{c}{Base} &                      \multicolumn{1}{c}{Mit} &                      \multicolumn{1}{c}{Base} &                      \multicolumn{1}{c}{Mit} \\
\midrule
Ekstrand et al. & AvgRating &         0.905 &         0.914 &      {\scriptsize \^{}}-0.032 &         {\scriptsize *}-0.027 &  {\scriptsize \^{}}0.047 &  {\scriptsize \^{}}0.045 &         1.239 &         1.246 &                    0.025 &                    0.024 &  {\scriptsize \^{}}0.060 &  {\scriptsize \^{}}0.070 \\
          & FunkSVD &         0.881 &         0.894 &  {\scriptsize \^{}}-0.032 &     \bftab      -0.023 &  {\scriptsize \^{}}0.052 &  {\scriptsize \^{}}0.051 &         1.255 &         1.268 &         {\scriptsize *}0.039 &                        0.039 &  {\scriptsize \^{}}0.040 &  {\scriptsize \^{}}0.052 \\
          & ItemKNN &   0.865 &   0.882 &  {\scriptsize \^{}}-0.034 &  {\scriptsize *}-0.026 &  {\scriptsize \^{}}0.055 &  {\scriptsize \^{}}0.056 &   1.218 &   1.230 &   {\scriptsize *}0.037 &   {\scriptsize *}0.035 &  {\scriptsize \^{}}0.064 &  {\scriptsize \^{}}0.072 \\
          & UserKNN &         0.896 &         0.911 &  {\scriptsize \^{}}-0.035 &                 -0.025 &  {\scriptsize \^{}}0.056 &  {\scriptsize \^{}}0.058 &         1.226 &         1.239 &      {\scriptsize \^{}}0.047 &         {\scriptsize *}0.054 & \bftab {\scriptsize \^{}}0.036 &  {\scriptsize \^{}}0.045 \\
Kamishima et al. & PMF BDist &   \bftab 0.863 &   0.870 & \bftab {\scriptsize \^{}}-0.029 &  {\scriptsize \^{}}-0.046 &  {\scriptsize \^{}}0.056 & \bftab {\scriptsize \^{}}0.032 &   1.172 &   1.179 &  \bftab 0.014 &         {\scriptsize *}0.029 &  {\scriptsize \^{}}0.067 &  {\scriptsize \^{}}0.029 \\
                 & PMF Mean & \bftab  0.863 &   0.870 & \bftab {\scriptsize \^{}}-0.029 &  {\scriptsize \^{}}-0.048 &  {\scriptsize \^{}}0.056 &  {\scriptsize \^{}}0.056 &   1.172 &   1.179 &  \bftab 0.014 &   {\scriptsize *}0.025 &  {\scriptsize \^{}}0.067 &  {\scriptsize \^{}}0.054 \\
                 & PMF Mi & \bftab  0.863 &   0.870 & \bftab {\scriptsize \^{}}-0.029 &  {\scriptsize \^{}}-0.046 &  {\scriptsize \^{}}0.056 & \bftab {\scriptsize \^{}}0.032 &   1.172 &   1.179 & \bftab  0.014 &         {\scriptsize *}0.029 &  {\scriptsize \^{}}0.067 &  {\scriptsize \^{}}0.029 \\
Rastegarpanah et al. & ALS &        0.894 &   0.890 &  {\scriptsize \^{}}-0.034 &  {\scriptsize \^{}}-0.034 & \bftab {\scriptsize \^{}}0.035 &  {\scriptsize \^{}}0.033 &   1.490 &   1.189 &  {\scriptsize \^{}}0.145 &   0.029 & \bftab {\scriptsize \^{}}0.036 &  {\scriptsize \^{}}0.114 \\
Ashokan \& Haas & ALS Par &         0.867 &         0.868 &  {\scriptsize \^{}}-0.030 &  {\scriptsize \^{}}-0.029 &  {\scriptsize \^{}}0.056 &  {\scriptsize \^{}}0.034 & \bftab 1.145 & \bftab 1.146 &          0.016 &      \bftab    0.018 &  {\scriptsize \^{}}0.047 &  \bftab {\scriptsize *}0.017 \\
          & ALS Val &         0.867 &  0.867 &  {\scriptsize \^{}}-0.030 &  {\scriptsize \^{}}-0.030 &  {\scriptsize \^{}}0.056 &  {\scriptsize \^{}}0.057 & \bftab 1.145 &         1.150 &           0.016 &       \bftab    0.018 &  {\scriptsize \^{}}0.047 &      {\scriptsize \^{}}0.050 \\
          & ItemKNN Par &  0.865 &         0.866 &  {\scriptsize \^{}}-0.034 &  {\scriptsize \^{}}-0.033 &  {\scriptsize \^{}}0.055 &  {\scriptsize \^{}}0.036 &       1.176 &         1.183 &  {\scriptsize *}0.033 &  {\scriptsize *}0.045 &  {\scriptsize \^{}}0.061 &      {\scriptsize \^{}}0.058 \\
          & ItemKNN Val &    0.865 & \bftab  0.865 &  {\scriptsize \^{}}-0.034 &  {\scriptsize \^{}}-0.034 &  {\scriptsize \^{}}0.055 &  {\scriptsize \^{}}0.052    &  1.176 &         1.173 &  {\scriptsize *}0.033 &  {\scriptsize *}0.036 &  {\scriptsize \^{}}0.061 &      {\scriptsize \^{}}0.046 \\
Wu et al. & FairGo GCN &  0.895 &  0.892 &  {\scriptsize \^{}}-0.038 &  {\scriptsize \^{}}-0.034 &  {\scriptsize \^{}}0.048 &  {\scriptsize \^{}}0.045 &  1.609 &  1.283 &  {\scriptsize \^{}}0.151 &   0.038 &  {\scriptsize \^{}}0.113 &  {\scriptsize \^{}}0.113 \\
\bottomrule
\end{tabular}
}
\end{table}

\begin{table}[!t]
\vspace{-6mm}
\centering
\caption{\footnotesize Rating prediction (RP) considering age groups.}
\label{tab:rmse_age}
\resizebox{1.0\textwidth}{!}{
\begin{tabular}{ll|rrrrrr|rrrrrr}
\toprule
          &            & \multicolumn{6}{c|}{ML 1M} & \multicolumn{6}{c}{LFM 1K} \\
          &            & \multicolumn{2}{c}{RMSE $\downarrow$} & \multicolumn{2}{c}{DP $\downarrow$} & \multicolumn{2}{c|}{KS $\downarrow$} & \multicolumn{2}{c}{RMSE $\downarrow$} & \multicolumn{2}{c}{DP $\downarrow$} & \multicolumn{2}{c}{KS $\downarrow$} \\
          Paper & Model &           \multicolumn{1}{c}{Base} &           \multicolumn{1}{c}{Mit} &                      \multicolumn{1}{c}{Base} &                      \multicolumn{1}{c}{Mit} &                      \multicolumn{1}{c}{Base} &                      \multicolumn{1}{c|}{Mit} &           \multicolumn{1}{c}{Base} &           \multicolumn{1}{c}{Mit} &                   \multicolumn{1}{c}{Base} &                      \multicolumn{1}{c}{Mit} &                      \multicolumn{1}{c}{Base} &                      \multicolumn{1}{c}{Mit} \\
\midrule
Ekstrand et al. & AvgRating &         0.905 &         0.904 &  {\scriptsize \^{}}0.051 &  {\scriptsize \^{}}0.056 &  {\scriptsize \^{}}0.071 &  {\scriptsize \^{}}0.072 &         1.239 &         1.248 &                 0.040 &                    0.048 &  {\scriptsize \^{}}0.080 &  {\scriptsize \^{}}0.092 \\
          & FunkSVD &         0.881 &         0.886 &  {\scriptsize \^{}}0.042 &  {\scriptsize \^{}}0.045 &  {\scriptsize \^{}}0.073 &  {\scriptsize \^{}}0.081 &         1.255 &         1.264 &         0.032 &         0.035 &  {\scriptsize \^{}}0.083 &  {\scriptsize \^{}}0.086 \\
          & ItemKNN &  0.865 &   0.875 &  {\scriptsize \^{}}0.039 &  {\scriptsize \^{}}0.042 &  {\scriptsize \^{}}0.074 &  {\scriptsize \^{}}0.079 &   1.218 &   1.226 &  \bftab 0.019 & \bftab  0.028 &  {\scriptsize \^{}}0.088 &  {\scriptsize \^{}}0.092 \\
          & UserKNN &         0.896 &         0.902 &  {\scriptsize \^{}}0.047 &  {\scriptsize \^{}}0.050 &  {\scriptsize \^{}}0.092 &  {\scriptsize \^{}}0.103 &         1.226 &         1.233 &         0.034 &         0.031 &  {\scriptsize \^{}}0.087 &  {\scriptsize \^{}}0.095 \\
Kamishima et al. & PMF BDist & \bftab  0.863 &   0.872 &  {\scriptsize \^{}}0.039 &  {\scriptsize \^{}}0.031 &  {\scriptsize \^{}}0.084 & \bftab {\scriptsize \^{}}0.018 &   1.172 &   1.183 &   0.045 &  {\scriptsize \^{}}0.065 &  {\scriptsize \^{}}0.124 &  {\scriptsize \^{}}0.047 \\
                 & PMF Mean &  \bftab 0.863 &   0.872 &  {\scriptsize \^{}}0.039 & \bftab {\scriptsize \^{}}0.027 &  {\scriptsize \^{}}0.084 &  {\scriptsize \^{}}0.045 &   1.172 &         1.184 &   0.045 &  {\scriptsize \^{}}0.069 &  {\scriptsize \^{}}0.124 &  {\scriptsize \^{}}0.042 \\
                 & PMF Mi & \bftab  0.863 &   0.872 &  {\scriptsize \^{}}0.039 &  {\scriptsize \^{}}0.031 &  {\scriptsize \^{}}0.084 & \bftab {\scriptsize \^{}}0.018 &   1.172 &   1.183 &   0.045 &  {\scriptsize \^{}}0.064 &  {\scriptsize \^{}}0.124 &  {\scriptsize \^{}}0.047 \\
Rastegarpanah et al. & ALS &         0.894 &   0.892 & \bftab {\scriptsize \^{}}0.034 &  {\scriptsize \^{}}0.040 & \bftab {\scriptsize \^{}}0.034 &  {\scriptsize \^{}}0.037 &   1.490 &   1.185 &   0.033 &   {\scriptsize *}0.052 & \bftab {\scriptsize \^{}}0.017 &  {\scriptsize \^{}}0.064 \\
Ashokan \& Haas & ALS Par &        0.867 &         0.871 &  {\scriptsize \^{}}0.041 &  {\scriptsize \^{}}0.048 &  {\scriptsize \^{}}0.074 &  {\scriptsize \^{}}0.026 & \bftab 1.145 & \bftab  1.146 &         0.043 &  {\scriptsize *}0.046 &  {\scriptsize \^{}}0.082 & \bftab  {\scriptsize *}0.015 \\
          & ALS Val &         0.867 &   0.866 &  {\scriptsize \^{}}0.041 &  {\scriptsize \^{}}0.042 &  {\scriptsize \^{}}0.074 &  {\scriptsize \^{}}0.079 & \bftab 1.145 &         1.149 &         0.043 &  {\scriptsize *}0.046 &  {\scriptsize \^{}}0.082 &      {\scriptsize \^{}}0.077 \\
          & ItemKNN Par &   0.865 &         0.870 &  {\scriptsize \^{}}0.040 &  {\scriptsize \^{}}0.048 &  {\scriptsize \^{}}0.074 &  {\scriptsize \^{}}0.031 &      1.176 &         1.177 &   0.029 &           0.031 &  {\scriptsize \^{}}0.085 &      {\scriptsize \^{}}0.029 \\
          & ItemKNN Val &   0.865 &  \bftab 0.864 &  {\scriptsize \^{}}0.040 &  {\scriptsize \^{}}0.042 &  {\scriptsize \^{}}0.074 &  {\scriptsize \^{}}0.071 &       1.176 &         1.172 &   0.029 &                 0.032 &  {\scriptsize \^{}}0.085 &      {\scriptsize \^{}}0.083 \\
Wu et al. & FairGo GCN &  0.895 &   0.908 &  {\scriptsize \^{}}0.040 &  {\scriptsize \^{}}0.044 &  {\scriptsize \^{}}0.070 &  {\scriptsize \^{}}0.074 &  1.609 &  1.277 &  0.043 &  {\scriptsize *}0.056 &  {\scriptsize \^{}}0.079 &  {\scriptsize \^{}}0.120 \\
\bottomrule
\end{tabular}
}
\vspace{-5mm}
\end{table}

\subsection{Impact on Recommendation Utility (RQ1)}
In a first analysis, we assess the impact of mitigation on recommendation utility, focusing on the NDCG/RMSE columns provided in the aforementioned tables.

In a TR task, we observed that the NDCG achieved by the untreated models (Base) in ML 1M was in the range $[0.110,0.140]$, except for SLIM-U, FunkSVD, LBM, and STAMP, whose NDCG was lower ($\le 0.084$). Mitigating unfairness (Mit) in ML 1M did not generally result in a substantial change in utility ($\pm 0.006$ gender; $\pm 0.003$ age). Higher changes were observed in two cases: SLIM-U treated with Burke et al.'s mitigation (stable for gender; $-0.036$ age) and LBM treated with Frisch et al.'s ($-0.023$ gender; stable for age). In LFM 1K, the untreated models (Base) got an NDCG in $[0.204,0.406]$, overall higher than ML 1M. The models ranking based on NDCG differs for several models from ML 1M. Though their utility was relatively high, PMF, FunkSVD, LBM, and STAMP were still under-performing in LFM 1K. The treated models (Mit) showed changes in NDCG ($\pm 0.009$ gender; $\pm 0.018$ age) larger in magnitude than ML 1M. SLIM-U with Burke et al.'s mitigation ($-0.047$ gender; $-0.141$ age) and LBM with Frisch et al.'s mitigation ($+0.068$ gender; $+0.069$ age) led to higher changes in NDCG.       

Considering an RP task, the untreated models (Base) achieved an RMSE in the range $[0.863,0.905]$ in ML 1M. By mitigating (Mit) in ML 1M, no substantial changes were observed ($\pm 0.017$ gender; $\pm 0.013$ age). In LFM 1K, the untreated models (Base) achieved a higher RMSE, in the range $[1.145,1.255]$. ALS and GCN are the lowest performers ($1.490$ and $1.609$, respectively). The treated models (Mit) showed minimal ($\pm 0.0135$ gender; $\pm 0.012$ age) which are similar to the changes in ML 1M. ALS under Rastegarpanah et al.'s mitigation lowered RMSE ($-0.301$ gender; $-0.305$ age), as well as GCN under Wu et al.'s mitigation ($-0.326$ gender; $-0.332$ age). 

\vspace{-2mm}
\hlbox{Observation 1}{In general, the mitigation procedures did not substantially impact on recommendation utility, regardless of the sensitive attribute, data set, task. The impact is larger in LFM 1K than ML 1M. }
\vspace{-7mm}

\subsection{Impact on Group Unfairness (RQ2)}
In a second analysis, we investigated the impact of mitigation on unfairness.  For each table and data set, we consider the DP and KS columns.

We start from a TR task, focusing our presentation on the subset of models that achieved a reasonable NDCG ($\le 0.110$ for ML 1M; $\le 204$ for LFM 1K). In ML 1M, the DP and KS achieved by the untreated models (Base) laid in the ranges ($[0.016, 0.038]$ gender; $[0.018, 0038]$ age) and ($[0.007, 0.074]$ gender; $[0.006, 0.060]$ age), respectively. Without any mitigation, in terms of DP, BiasedMF, NCF, and PMF ($\le 0.020$ gender; $\le 0.027$ age) were fairer than TopPopular, UserKNN, and ItemKNN ($\ge 0.031$ gender; $\ge 0.027$ age). To some surprise, when KS was considered, we observed a different pattern. TopPopular was the fairest model ($0.007$ gender; $0.006$ age), followed by NCF and PMF ($0.024$ and $0.026$ gender; $0.031$ and $0.027$ age), ItemKNN and BiasedMF ($0.030$ and $0.033$ gender; $0.029$ and $0.042$ age), and UserKNN ($0.074$ gender; $0.060$ age). By mitigating (Mit), DP went down to the range ($[0.013, 0.030]$ gender; $[0.017, 0.034]$ age), while KS laid in the range ($[0.006, 0.052]$ gender; $[0.006, 0.051]$ age). In LFM 1K,  models were less fair than in ML 1M. The untreated models (Base) achieved a DP in the ranges ($[-0.046, -0.127]$ gender; $[0.010, -0.044]$ age) and a KS in the ranges ($[0.001, 0.067]$ gender; $[0.006, 0.133]$ age). The models ranking in terms of DP and KS was similar between LFM 1K and ML 1M. Once mitigated (Mit), interestingly, we observed that re-sampling by Ekstrand et al. resulted in a decrease of fairness for TopPopular in terms of DP on gender groups ($0.017$), and for TopPopular, ItemKNN and UserKNN on age groups ($\ge 0.06$). These findings are replicated for ItemKNN in terms of KS on gender groups ($0.03$), while, for age groups KS was substantially lowered ($0.015$). Other cases did not lead to substantial changes. 

In a RP task, in ML 1M, untreated models (Base) achieved a DP in $[-0.038, \\ -0.025]$ (gender) and $[0.034, 0.051]$ (age), and a KS in $[0.035, 0.056]$ (gender) and $[0.034, 0.092]$ (age). With no mitigation, there were minimal differences in terms of DP between models for the attribute gender (avg. $0.033$, std. dev. $0.003$). For the attribute age, the untreated models had similar DP (avg. $0.041$, std. dev. $0.005$). Considering KS, comparable estimates across models were observed (avg. $0.053$, std. dev. $0.003$ gender; avg. $0.076$, std. dev. $0.007$ age). ALS ($0.035$ gender; $0.034$ age) resulted in fairer outcomes in terms of KS. Treated models (Mit) showed stable fairness ($\pm 0.010$ gender; $\pm 0.008$ age) in all cases, except for Kamishima et al. ($\pm 0.019$ gender; $\pm 0.012$ age) when DP was considered. In terms of KS, models treated with Kamishima et al.'s mitigation (for gender only PMF BDist and PMF Mi) and Ashokan et al.'s mitigation (parity setting) were substantially fairer ($\ge 0.019$ gender; $\ge 0.039$ age), while other treated models did not benefit from the mitigation ($\pm 0.003$ gender; $\pm 0.011$ age). In LFM 1K, untreated models (Base) achieved a DP in $[0.014, 0.151]$ (gender) and $[0.019, 0.045]$ (age), and a KS in $[0.036, 0.113]$ (gender) and $[0.017, 0.124]$ (age). Without mitigating, findings in ML 1M held in LFM 1K, except for the high DP ($0.151$) and KS ($0.113$) of GCN for gender. Treated models (Mit) instead showed stable fairness ($\le 0.015$ gender; $\le 0.009$ age) except for Kamishima et al. ($\ge 0.019$ age), ALS ($0.116$ gender; $0.019$ age), GCN ($0.113$ gender; $0.013$ age), in terms of DP (opposite to ML 1M). In terms of KS, except the mitigations of Kamishima et al. and Ashokan et al. (parity), treated models did not benefit from mitigation ($\le 0.015$ gender; $\le 0.005$ age).

\vspace{-1.5mm}
\hlbox{Observation 2}{Unfairness depends on the mitigation, model, and fairness notion. Often the mitigation impact is small. Lowering DP does not imply lowering KS, and viceversa. Unfairness is higher in LFM than ML. }
\vspace{-7mm}

\subsection{Relationships between Representation and Unfairness (RQ3)}
In a third analysis, we analyzed whether the disparate impact always harms minority groups (see group representations in Table \ref{tab:repr_datasets}), based on the sign of DP.    

In a TR task, positive values of DP mean that models advantage the majority (majority group's NDCG higher than minority's group NDCG; the higher the NDCG the higher the utility). Conversely, negative values show an advantage for the minority. From our results, untreated models (Base) negatively impacted on the minority for both gender (in all cases significantly) and age (9/10 times significantly) in ML 1M. Though unfairness was reduced through mitigation, the same observations were still valid on treated models (Mit). To some surprise, the majority groups were negatively impacted for both attributes in LFM 1K (7/10 times significantly for gender and 2/10 times significantly for age) by untreated models (Base). By mitigating (Mit), 7 out of 10 treated models were significantly unfair for gender. For age, observations were similar.    

Considering a RP task, positive values of DP mean that models advantage the minority (majority group's RMSE higher than minority's group RMSE; the higher the RMSE the lower the utility). Conversely, negative values show an advantage for the majority. The results showed that the minority age group was advantaged in both data sets (in all cases significantly in ML 1M) by untreated models (Base). The minority group was also advantaged in LFM 1K for the gender attribute, significantly 7/13 times. Conversely, the majority gender group was advantaged in ML 1M, significantly in all cases. Similarly to the TR task, treated models (Mit) were still significantly unfair against the group disadvantaged in the untreated model. 

\vspace{-1mm}
\hlbox{Observation 3}{The disparate impact does not always harm the minority group. The latter was advantaged for both attributes in LFM 1K (TR), in both data sets for age and in LFM 1K for gender (RP).}
\vspace{-4.5mm}

\section{Discussion and Conclusions} \label{sec:discussion}
In this last section, we connect our findings and present the resulting conclusions.

\vspace{1mm} \noindent \textbf{Reproducibility}. Several challenges emerged while reproducing existing procedures. For instance, the code base modularity should be improved to easily accommodate different data sets as an input. Moreover, many procedures required extensive computational resources to treat the recommendation models. This issue prevented us from using larger data sets, e.g., LFM 360K, and questions scalability. Future works should account for modularity and efficiency.% and report computational time and memory analysis. 

\vspace{1mm} \noindent \textbf{Optimization}. Mitigating unfairness adds additional hyper-parameters and often requires to deal with a trade-off between recommendation utility and unfairness. It is hence challenging to tune the hyper-parameters. While we provide results for an arbitrary optimal setup, it is up to the stakeholders to select the trade-offs most suitable for their goals. One of the future directions should be to find novel mitigation procedures that embed the constraint on recommendation utility more strictly, to avoid convoluted decisions on the mentioned trade-off.  

\vspace{1mm} \noindent \textbf{Comparability}. Our study showed that there is an abundance of evaluation metrics to assess fairness and that, despite several papers using similar data sets (e.g., ML 1M), the evaluation setting was often different. Our paper shows the first attempt of comparing a wide range of mitigation procedures under the same evaluation protocol, considering two relevant yet transferable fairness notions. Despite the common protocol, we however could not conclude whether a mitigation is better than another in a given context, given that many of them could not be easily transferred across models. In the future, a mitigation procedure should be tested across recommendation models, data sets and sensitive attributes.  

\vspace{1mm} \noindent \textbf{Impact}. Our results showed that the impact of the mitigation procedure on utility is often negligible. However, depending on the recommendation model, the data set, and the task, mitigation procedures do not always substantially reduce unfairness. Moreover, being fair in terms of independence does not imply higher fairness in terms of equity. Future work should study the friction across fairness notions, and ensure that the unfairness reduction achieved offline can then provide tangible impacts when moved online.    

\vspace{1mm} Overall, our analyses showed that reproducing research in this area is still a challenging task hence and call for more rigor and shared practices in this area. Motivated by our findings, we will extend our analyses to papers published in other outlets and to other notions of consumer fairness. We also plan to devise novel mitigation procedures, following the lessons learnt from this study (e.g., modularity, efficiency, optimization, comparability, impact).  

\bibliographystyle{splncs04}
\bibliography{bibliography.bib}

\end{document}